\documentstyle[11pt,epsfig]{article}

\newcommand{\sect}[1]{\setcounter{equation}{0}\section{#1}}

\def\V{{\cal V}}
\def\a{\alpha}
\def\b{\beta}
\def\g{\gamma}
\def\l{\lambda}
\def\La{\Lambda}
\def\s{\sigma}

\def\Om{\Omega}
\def\t{\theta}
\def\Th{\Theta}
\def\o{\over}

\def\mbf#1{\mbox{\boldmath$#1$}}		

\def\VEV#1{\left\langle #1 \right\rangle}   
\def\frac#1#2{{\textstyle{ #1 \over #2 }}}  
\def\til{\tilde}
                  
\def\bra#1{\left\langle #1\right|}          
\def\ket#1{\left| #1\right\rangle}          
\def\dif#1{{\rm d}#1}
\def\der#1#2{{\dif#1\over\dif#2}}

\def\da{\dagger}                                 
\def\ZZ{{\rm Z \!\! Z}}

\begin{document}
 
\topmargin 0pt
\oddsidemargin 5mm

\newpage
\begin{titlepage}
\vspace{2cm}
\begin{center}
{\bf{{\Large A LOCAL AND INTEGRABLE }}} \\
{\bf{{\Large LATTICE REGULARIZATION }}} \\
{\bf{{\Large OF THE MASSIVE THIRRING MODEL }}} \\
\vspace{2cm}
{\large C.Destri }\footnote[1]{E-MAIL: Destri@mi.infn.it} \\
{\em Dipartimento di Fisica, Universit\`{a} di Milano and} \\
{\em INFN, Sezione di Milano, I-20133 Milano, Italy}\\
\vspace{2cm}
{\large T.Segalini }\footnote[4]{E-MAIL: Segalini@.pr.infn.it} \\
{\em Dipartimento di Fisica, Universit\`{a} di Parma and} \\
{\em INFN, Gruppo Collegato di Parma, I-43100 Parma, Italy}\\
\end{center}

\vspace{2cm} \centerline{{\bf Abstract}} 

The light--cone lattice approach to the massive Thirring model is
reformulated using a local and integrable lattice Hamiltonian written
in terms of discrete fermi fields. Several subtle points concerning
boundary conditions, normal--ordering, continuum limit, finite
renormalizations and decoupling of fermion doublers are
elucidated. The relations connecting the six--vertex anisotropy and
the various coupling constants of the continuum are analyzed in
detail.

\vspace{.5cm}
 
\vfill
\noindent
IFUM--510--FT \hfill \\
\noindent
UPRF--95--424 \hfill \\
\noindent
hep--th/9506120 \hfill {May 1995}
\end{titlepage}

\sect{Introduction}

A very convenient way to non--perturbatively regularize a QFT is to
put the dynamical variables of the theory on a regular spacetime
lattice (in the functional--integral formulation) or on a regular
space lattice (in the hamiltonian framework). This introduces a
``natural'' cutoff, roughtly equal to the inverse of the lattice
spacing, either on both energy and momentum or on momentum
alone. Usually this procedure breaks the symmetry properties of the
action down to a lower level: Lorentz or Euclidean invariance reduces
to invariance under discrete subgroups, scale invariance in massless
theory is broken explicitly by the cutoff, and very often also
internal symmetries, either global or local, are difficult to keep.

Therefore it is very interesting to find regularization procedures
that preserve as much as possible of the characteristics of the
continuum theory. This issue is particularly important in the case of
two--dimensional models which are integrable at tree level and are
supposed to be so also at the full quantum level. One would like to
have a non--perturbative lattice definition of such quantum theories
which preserves integrability. 

A quite general solution to this problem is based on the so--called
light--cone approach \cite{dv}, in which the 2D Minkowski spacetime is
discretized in light--cone coordinates. The basic object in this
approach is the $R$--matrix, that is a solution of the Yang--Baxter
equations which characterize the factorized scattering of a 2D
integrable QFT. This $R$--matrix is regarded as a collection of
quantum amplitudes for the scattering of ``bare'' objects, which move
with the rapidity cutoff $\Th$, on each vertex of the light--cone
lattice, casting the model in question in the form of a vertex
model. Then the full machinery based on monodromy and transfer
matrices \cite{russi}\cite{baxter} \cite{devega}  
can be set up and the algebrized or analityc Bethe
ansatz (BA) can be used to completely diagonalize the transfer matrix and,
with it, the total momentum, the Hamiltonian and all other conserved
charges. The continuum limit may then be explicitly performed by
letting $\Th$ go to infinity in a well defined way as the lattice
spacing vanishes.

A drawback of the standard light--cone approach is the nonlocality of
the lattice Hamiltonian. While this does not constitute a real problem
for the continuum limit, either at the bare or renormalized level, it
makes more difficult to properly handle the full excitation spectrum
and to study the conformal limit, which allows to identify the
integrable model at hand as a perturbed CFT. A sligthly modified
version of the light--cone approach without such difficulties was
recently put forward in \cite{rs}: rather than as logarithm of the
unit time evolution operator (or diagonal--to--diagonal transfer
matrix), the lattice Hamiltonian is identified as the first of the
series of local charges obtained by suitably differentiating the
alternating transfer matrix with respect to the spectral
parameter. Such identification was made before, whithin a different
context, in \cite{dvwoi}. The basic property of this modified approach is
the locality of the lattice Hamiltonian, which allows to safely regard 
the time as continuous while the space is still discrete, restricting
the UV cutoff only to the space momentum. 

In this work we present a detailed application of the local
light--cone approach to the massive Thirring model. This is probably
the simplest case, being based on the well known, almost paradigmatic
six--vertex $R$--matrix, without any quantum group restriction, and
was for such reason the first model studied also in the nonlocal
approach \cite{dv1}. Nonetheless there are some interesting
non--trivial points that require a careful examination.

First of all one must take into account the Nielsen--Ninomiya theorem
\cite{nini}, since the lattice Hamiltonian is local and
chiral--invariant in the $\Th\to\infty$ limit (this is one of the most
important differences of the light--cone approach with respect to 
L\"uscher's regularization based on the XYZ spin chain \cite{lush}: the
latter is indeed integrable but has neither $U(1)$ invariance nor a
local implementation of chiral transformations). As a consequence one
finds the ``fermion doublers'' both in the perturbative spectrum and
in the exact Bethe ansatz spectrum. It is then important to check
whether these massless doublers indeed decouple from the massive
Thirring particles.  We show the answer to be affirmative even off
shell, for the local continuum fields, although the mechanism is quite
non--trivial.

Secondarily, we examine in detail the problem of boundary conditions
and their effects on the exact spectrum. In particular, by carefully 
handling a completely fermionic formulation we are able to show that 
the excitations over the ground state carry the correct $U(1)$ charge
which corresponds to dressed fermions interpolated by the bare fields.
This should be compared with the result proper of the periodic spin chain,
with excitations carrying half the $U(1)$ charge of the fermions. 

Another interesting point concerns the structure of the perturbative
vacuum on the lattice: while the one--particle spectrum over the
emptied Dirac sea (the state killed by the local fermi fields) has a
anisotropy--dependent zeroes and no simmetry between positive and
negative energies, this simmetry is restored and the anomalous zeroes
move to the boundary of the Brillouin zone simply by normal--ordering
the $U(1)$ currents in the lattice Hamiltonian. This facts allows to 
isolate the effects of the interaction, even before the continuum
limit, in a cutoff--dependent mass renormalization and
in a finite rescaling of the velocity of light.

The finite renormalization of the speed of light is one last subtlety
that requires a proper treatment. While such renormalization is absent
in the nonlocal light--cone approach, where the simmetry between space
and time is mantained all along, nothing forbids it in the local
formulation, since time may be regarded as already continuous while
space is still discrete. We handle this by intruducing a time unit
$a_t$ which is independent from the lattice spacing $a$ of the space
chain. The velocity of light, either bare or renormalized, emerges
quite naturally as finite ratio $a/a_t$.

This paper is organized as follows. In section \ref{sec:basic} we
describe the basic framework of vertex models and derive in a purely
algebraic way the local lattice Hamiltonian, using first the
$R$--matrix written in spin language.  In section \ref{sec:fermi} we
discuss the subleties related to the formulation on the ligh--cone
lattice of the system using a fermionic approach. Indeed the
translation of the $R$--matrix in fermionic variable is quick (after
taking in account some important changes of sign due to the Fermi
statistic), but involves a careful definition of the boundary
condition.  The explicit form of the hamiltonian and the dispersion
laws for the lattice fermions derived in this fermionic setup are
shown in section \ref{sec:form}, where it is also discussed the
normal--ordering prescription we adopt for the $U(1)$ currents over a
completely occupied Dirac sea. This corresponds to an antiferromagnetic
ground state in spin language. The continuum limit is considered in
section \ref{sec:continuum}, where abelian bosonization tricks are
used to disentangle the mixed currents terms that arise in the naive
continuum limit. In this way we shows that in the continuum
Hamiltonian does describe two fermi fields, one massless and one
massive. In section \ref{sec:BA} the results of the Bethe ansatz are
briefly rewieved, showing some novelty regarding the meaning of the
hole charge in the framework with antiperiodic boundary conditions and
the matching between the dispersion laws perturbatively derived from
the lattice Hamiltonian and the exact one based on the Bethe
ansatz. Finally, in \ref{sec:gg'b}, we study the effects of
renormalization and of the trasformation to the decoupled description
on the relation between the various coupling constants: for instance,
the current--current Thirring coupling and the sine--Gordon coupling
constant $\b$ are related in the standard one only after a suitable
power serie redefinition. Some comments on the results obtained and on
possible further developments can be found in \ref{sec:fine}.

\sect{The basic framework}\label{sec:basic}

It is well known \cite{baxter} \cite{devega} that the 6V model, as
well as the XXZ spin chain related to it, may be formulated starting
from a collection of two--dimensional vector spaces
$\{\V_j,\,j=1,2\ldots,N\}$ and local R-matrices $R_{ij}$ acting on the
tensorial product $\V_i\otimes\V_j$ of two such spaces.  These
$R_{ij}$ are written in terms of the Pauli's matrices
$\sigma^x_j,\,\sigma^y_j,\,\sigma^z_j$, $j=1,2,\ldots,N$, as
\begin{equation}\label{eq:rsigma}
	R_{ij}(\l)={{1+c}\over2}+
		{{1-c}\over2} \sigma_i^z\sigma_j^z +
		b \left[
		\sigma_i^+\sigma_j^-+ \sigma_i^-\sigma_j^+\right]
\end{equation}
where the $\sigma^\pm=\sigma^x \pm i\sigma^y$ and the
trigonometric Boltzmann weights $b,c$ 
are parametrized as follows by the spectral parameter $\l$:
\begin{eqnarray}\label{eq:pesi}
	b=b(\l)\equiv {{\sinh{\l}}\over{\sinh{(i\gamma-\l)}}}
			\nonumber\\                
	c=c(\l)\equiv {{i \sin{\gamma}}\over{\sinh{(i\gamma-\l)}}}
\end{eqnarray}
The choice of weights made here guarantees that the $R$--matrices are
unitary for real $\l$ and $\g$, that is $R_{jk}^\da R_{jk}=1$. It is
straightforward to check that this reduces to the identities
$|b|^2+|c|^2=1$ and ${\bar b}c+b{\bar c}=0$. As we shall see below,
the unitarity property is important in order to interpret the transfer
matrix as a temporal evolution operator. The regularity condition
of the R-matrix is fulfilled by eq.(\ref{eq:rsigma}) as
$R_{jk}(0)=1$. Most importantly, the $R$--matrices satisfy the
Yang-Baxter equations (YBE)\cite{ji}
\begin{equation}\label{eq:ybe}
	R_{ij}(\l)R_{jk}(\l+\mu)R_{ij}(\mu) 
	= R_{jk}(\mu)R_{ij}(\l+\mu)R_{jk}(\l)
\end{equation}
which ensure the integrability of the 6V model in any framework.

The `bare scattering' $S$--matrices are defined as
\begin{equation}\label{eq:smatrix}
	S_{ij}(\l)=P_{ij}R_{ij}(\l)
\end{equation}
where the permutation operators $P_{ij}$ interchange
the vector space $\V_i$ and $\V_j$:
$P_{ij}\V_i\otimes\V_j=\V_j\otimes\V_i$. In terms of the $S$--matrices
the YBE  eq.(\ref{eq:ybe}) can be reformulated as
\begin{equation}\label{eq:ybes}
	S_{jk}(\l)S_{ik}(\l+\mu)S_{ij}(\mu)
	=S_{ij}(\mu)S_{ik}(\l+\mu)S_{jk}(\l)
\end{equation}

Let's now introduce the fully inhomogeneous monodromy matrix
$T(\l|\{\t_i\})$ associated with the auxiliary ``horizontal'' vector
space $\V_0$
\begin{equation}\label{eq:mono}
	T(\l|\{\t_i\})=S_{10}(\l+\t_1)S_{20}(\l+\t_2)\ldots
	S_{N0}(\l+\t_N) \equiv \pmatrix{A&B\cr C&D\cr}
\end{equation} 
where the operators $A,\,B,\,C,\,D$ act in the full Hilbert space
$\V_1\otimes\V_2\ldots\otimes\V_N$. The monodromy matrix, thanks to
the YBE, satisfies the Yang--Baxter algebra (YBA)
\begin{equation}\label{eq:ybalg}
	R(\l-\mu)\left[T(\l|\,\{\t_i\}) \otimes
	T(\mu|\,\{\t_i\}) \right] =\left[T(\mu|\,\{\t_i\}) 
	\otimes T(\l|\,\{\t_i\}) \right]R(\l-\mu) \;.
\end{equation}
These implies a set of commutation rules for $A,\,B,\,C,\,D$, among
which the following play a central r\^ole in the algebraic Bethe ansatz: 
\begin{eqnarray}\label{eq:ybalgI}
	b(\mu-\l)A(\l)B(\mu) \!\!&=&\!\! 
	+B(\mu)A(\l) - c(\mu-\l)B(\l)A(\mu)	\nonumber \\
	b(\l-\mu)D(\l)B(\mu) \!\!&=&\!\! 
	+B(\mu)D(\l) - c(\l-\mu)B(\l)D(\mu)	         \\
	B(\l)B(\mu) \!\!&=&\!\! B(\mu) B(\l)  	\nonumber \;.
\end{eqnarray}
Taking the trace of the monodromy matrix over the horizontal space we obtain
the transfer matrix
\begin{equation}\label{eq:trasfmat}
	t(\l|\{\t_i\})= \mathrm{tr}_0 T(\l|\{\t_i\})
\end{equation}
For fixed arbitrary set of vertical 
inhomogeneities $\{\t_i\}$, thanks again to the YBE, the
transfer matrices form an infinite set of commuting operators.
\[
		[t(\l|\{\t_i\})\,,\,t(\mu|\{\t_i\})]=0
\]
Since we are trying to regolarize a relativistic QFT on a light-cone
lattice, we choose the vertical inhomogeneities in a particular way,
consistent with the propagation of `bare particles' moving along the
two diagonal directions with cutoff rapidity $\pm \Th$, respectively:
\begin{equation}\label{eq:altomo}
	\t_i=(-1)^{i+1}\Th \;,\quad i=1,2,\ldots,2N \;.
\end{equation} 
We have changed $N$ to $2N$ to ensure periodic boundary conditions.

Inserting these alternating inhomogeneities in eq.(\ref{eq:mono}) 
we obtain the {\em alternating} monodromy matrix 
\begin{equation}\label{eq:altmono}
	T(\l|\,\Th)= S_{10}(\l+\Th)S_{20}(\l-\Th)\ldots
                  S_{2N\,0}(\l-\Th)
\end{equation}
and, taking the trace as in eq.(\ref{eq:trasfmat}, the {\em alternating}
transfer matrix $t(\l|\,\Th)=\mathrm{tr}_0 T(\l|\,\Th)$.

The regularity condition $R_{jk}(0)=1$ and the permutation algebra 
\begin{equation}\label{eq:permut}
	P_{ij}A_{kn} = \cases{ A_{kn}P_{ij} &$i,j,k,n$ all distinct\cr
		A_{in}P_{ij} &$j=k;\;i,j,n$ all distinct\cr
		A_{ki}P_{ij} &$j=n;\;i,j,k$ all distinct\cr} \;,
\end{equation}
which holds for any operator $A_{ij}$ acting nontrivially only on
$\V_i\otimes\V_j$, imply the fundamental relation
\begin{equation}\label{eq:ul}
	t(\Th|\,\Th)  = U_L  \;,\quad t(-\Th|\,\Th) = U_R^\da  \;.
\end{equation}
Here $U_R$ and $U_L$ are the right and left diagonal
transfer matrices (they move by one lattice spacing in right--upward
$x+t$ and left--upward $x-t$ direction respectively) defined as
\begin{eqnarray}\label{eq:ulur}
	U_L = V R_{12}R_{34}\ldots R_{2N-1\,2N} \\
	U_R =V^{-1}R_{12}R_{34}\ldots R_{2N-1\,2N} 
\end{eqnarray} 
where $R_{jk}=R_{jk}(2\Th)$ and $V$ is the left shift operator 
$V=P_{1\,2N}P_{2\,2N}\ldots P_{2N-1\,2N}$. The derivation of these 
formulae is purely algebraic; for $U_L(\Th)$ we have
\begin{eqnarray}\label{eq:trictracI}
	t(\Th|\,\Th)\!&=&\! \mathrm{tr}_0\prod_{j=1}^N 
		S_{2j-1\,0}(2\Th) P_{2j\,0} 		\nonumber\\
	&=&\! \left( {\mathrm{tr}}_0\, P_{2N\,0} \right)
	\left[ \prod_{j=1}^{N-1} S_{2j-1\,2N}(2\Th) P_{2j\,2N}
	\right] 	S_{2N-1\,2N}(2\Th)		\nonumber\\
	&=&\! \left( \prod_{j=1}^{N-1} P_{2j\,2N} \right)
	\prod_{j=1}^N  P_{2j-1\,2j}R_{2j-1\,2j}(2\Th) 	\nonumber\\
	&=&\! V \prod_{j=1}^N R_{2j-1\,2j}(2\Th)   	\nonumber\\
	&=&\! U_L					
\end{eqnarray} 
with a similar calculation for $U_R$.  

The unit time evolution operator is ${\hat U}=U_RU_L$: it causes a
displacement $a_t$, the lattice spacing in the time direction, 
upwards on the light-cone lattice, leading to the following
definition of the lattice Hamiltonian:
\begin{equation}\label{eq:hnonlocal}
	{\hat H} = ia_t^{-1}\log {\hat U} \;.
\end{equation}
Evidently this Hamiltonian is nonlocal. Similarly nonlocal is the 
lattice momentum operator, naturally defined as 
\[
	P = -ia^{-1}\log V^2  \;,
\]
where $a$ is the lattice spacing in the space direction. 
On the other hand, from the commuting family of alternating transfer
matrices it is possible to obtain a full hierarchy of local charges in
involution. It suffices to take the logaritmic derivative of
$t(\l|\,\Th)$ with respect to the spectral parameter at $\l=\pm\Th$:
\begin{equation}\label{eq:charges}
	H^\pm_n = i^{-1-n}{{\partial^n}\over{\partial\l^n}}
	\log t(\l|\Th)\Bigl|_{\l=\pm\Th} \;.
\end{equation}
By purely algebraic calculations similar to those of
eq.(\ref{eq:trictracI}), one verifies that $H^\pm_n(\Th)$ couples $2n+1$
neighboring sites. Unlike in eq.(\ref{eq:trictracI}), in this derivation
it is crucial that the $R$--matrices satisfy the YBE. Since 
$H^\pm_n(\Th)$ commutes also with $U(\Th)$, it is a conserved charge.

The charges of level 1 read
\begin{equation}\label{eq:othercharges}
	H_1^+ = \sum_{j=1}^N h_{2j-1}(2\Th) \;,\quad
	H_1^- = \sum_{j=1}^N h_{2j}(-2\Th) 
\end{equation}
in terms of the `Hamiltonian density'
\begin{equation}\label{eq:otherchargesI}
	h_n(\l) = - R_{n\,n+1}(\l)^{-1} \left[{\dot R}_{n\,n+1}(\l) 
	+{\dot R}_{n-1\,n}(0) R_{n\,n+1}(\l) \right] \;.
\end{equation}
With them, one can now define the local Hamiltonian 
\begin{equation}\label{eq:localham}
	H = {1\over{2a_t}} \left[H^+_1 + H^-_1\right] \;,
\end{equation}
which is indeed hermitean thanks to the unitarity of $R-$matrix.  Of
course, with this choice of hamiltonian, the evolution operator is
$U(t)=e^{-itH}$, with the time $t$ continuous and $a_t$ merely fixing
the scale of time or energy.

\sect{Fermionic formulation}\label{sec:fermi}

The $U(1)$ invariance of the 6V $R-$matrix corresponds, in the
light--cone framework, to the conservation of bare particles.  In fact
the ferromagnetic state with all spins up, $\ket{++\ldots +}$, may be
regarded as `bare vacuum state' (the state with no bare
particles). Then we can say that a state with $r$ flipped spins
located at $1\le j_1\le j_2\ldots\le j_r\le 2N$, that is the state
\[
	\ket{j_1,j_2,\ldots,j_r} = 
	\s^-_{j_1} \s^-_{j_2}\ldots \s^-_{j_r} \ket{++\ldots +}
\]
contains exactly $r$ bare particles at the same locations. The
particle number $r$ is conserved in time, that is along the vertical
direction throughout the lattice, thanks to the $U(1)$ invariance of
the $R-$matrix.

It is clear that these particles are identical and satisfy the Pauli
exclusion principle, since $(\s^-_j)^2=0$. On the other hand, since
$[\s^-_j\,,\,\s^-_k]=0$ for $j\neq k$, they are of bosonic type. This
can be remedied by means of the well known Jordan--Wigner
transformation from the spin operators $\s^-_j$ and $\s^+_j$ to
lattice fermion fields
\begin{equation}\label{eq:JW}
	\psi_j = \s^+_j \prod_{n=1}^{j-1}\s^z_n \; \quad 
	\psi_j^\da = \s^-_j \prod_{n=1}^{j-1} \s^z_n  
\end{equation}
satisfying the canonical anticommutation rules
\begin{equation}\label{eq:acar}
	\{\psi_j\,,\,\psi_k\} =0= \{\psi_j^\da\,,\,\psi_k^\da\}
	\;\quad \{\psi_j\,,\,\psi_k^\da\}= \delta_{jk} \;.
\end{equation}
The string of $\s^z_n$ in eq.(\ref{eq:JW}) has a nontrivial effect only
on the boundary conditions. In fact it cancels completely out of all
local $R-$matrices with neighboring indices such $R_{j\,j+1}$ (with
$1\le j\le 2N-1$), which have the fermionic form
\begin{equation}\label{eq:fermirma}
	R_{j\,j+1} = 1 +bK_{j\,j+1} + (c-1)(Q_j-Q_{j+1})^2
\end{equation}
where
\[
	K_{ij} = \psi_i^\da\psi_j+\psi_j^\da\psi_i \;,\quad
	Q_j = \psi_j^\da\psi_j = \frac12(1-\s_j^z) \;.
\]
Thus the string of $\s^z_n$ would also drop out of the evolution
operator $U$ and of the local hamiltonian $H_1$, if it were not for
the periodic boundary conditions. The troblesome object is
$R_{2N\,1}(2\Th)$, which reads in terms of fermion operators
\begin{equation}\label{eq:ultiR}
	R_{2N\,1} = 1 - b\left[ \psi_{2N}^\da\psi_1(-)^F + 
	(-)^F\psi_1^\da\psi_{2N} \right] + (c-1)(Q_1-Q_{2N})^2
\end{equation}
where 
\[
	(-)^F \equiv \prod_{j=1}^{2N} \s_j^z = e^{i\pi Q}
\]
is the longest possible string, that is the fermion signature, and
$Q=\sum_jQ_j$ is the total bare particle number. Similarly, since the
left shift operator $V$ acts on the Pauli matrices as
\[
	V^\da \mbf\s_j V = \mbf\s_{j+1}
\]
it cannot shift exactly also the fermion fields. Rather we have
\begin{equation}\label{eq:badshift}
	V^\da \psi_j V = \s^+_{j+1} \prod_{n=2}^j\s^z_n =
	\psi_{j+1} e^{i\pi Q_1} \;,
\end{equation}
for $j=1,2,\ldots,2N-1$, and 
\[
	V^\da \psi_{2N} V \equiv \psi_{2N+1} = -\psi_1(-)^F \;.
\]
Together with eq.(\ref{eq:ultiR}), this last relation suggests that PBC
on the spin operators become a sort of F-twisted boundary conditions
\begin{equation}\label{eq:FTBC}
	\psi_{2N+1} \equiv -\psi_1 (-)^F
\end{equation}
on the fermion operators. However, if it is true that this guarantees
$R_{2N\,2N+1}=R_{2N\,1}$, eq.(\ref{eq:badshift}) prevents the
identification of $V$ with an exponential of the fermion total
momentum. In particular, $V^{2N}$ is the identity in the full vector
space $\V^{(2N)}$, and hence $V^{-2N}\psi_j V^{2N}=\psi_j$, which
shows the conflit between $\psi_{j+2N}\equiv V^{-2N}\psi_j V^{2N}$ and
the extension $\psi_{j+2N}=-\psi_j(-)^F$ of the F-twisted  relation
(\ref{eq:FTBC}) to all fermion operators. Of course, we could define
the true fermion shift operator $\til V$ through
\[
	{\til V}^\da \psi_j \til V = \psi_{j+1}
\]
for any $j$, and impose uniform F--twisted boundary conditions via 
\begin{equation}\label{eq:Ftbc}
	{\til V}^{-2N} \psi_j {\til V}^{2N} = -\psi_j(-)^F \;.
\end{equation}
Certainly $\til V$ commutes with $U$ and $H_1$ but, unlike them, it is
not related in any obvious way to the transfer matrix $t(\l|\,\Th)$,
which is the object we are able to actually diagonalize by means of
the algebraic BA.  Therefore the translation of the light--cone 6V
model and its BA solution from its original spin formulation into a
fermionic theory, by means of a straightforward application of the
Jordan--Wigner transformation, remains unsatisfactory due to
boundary effects.

Although we expect that these  boundary effects  will
loose importance in the limit $N\to\infty$, it is convenient to look
for a purely fermionic formulation, in which all basic objects, like
$R-$matrices and exchange operators, are written from the start in
term of fermion fields for any pair of indices.

To this end, let us notice that the matrices $R_{j\,j+1}(\l)$, whether
written in spin (eq.(\ref{eq:rsigma})) or fermion language
(eq.(\ref{eq:fermirma})), satisfy the YBE in the restricted form
\[
	R_{j-1\,j}(\l) R_{j\,j+1}(\l+\mu) R_{j-1\,j}(\mu) =
        R_{j\,j+1}(\mu) R_{j-1\,j}(\l+\mu) R_{j\,j+1}(\l) \;.
\]
But since $j-1$, $j$ and $j+1$ simply refer to three distinct
anticommuting fermions, the matrices
\[
	\til R_{ij}(\l) = 1 +b(\l)K_{ij} + [c(\l)-1](Q_i-Q_j)^2	
\]
will fulfill the general form (\ref{eq:ybe}) of the YBE, providing
another solution distinct from $R_{ij}(\l)$. In fact, 
$\til R_{ij}(\l)\neq R_{ij}(\l)$ for $|i-j|>1$.

Next we build the fermion permutation operators $\til P_{ij}$, defined
by the relations
\[
	\til P_{ij}\psi_i \til P_{ij}^{-1} = \psi_j 
	\;\quad \til P_{ij}=\til P_{ij}^{-1}=\til P_{ij}^\da  \,.
\]
They are written in terms of the fields simply as 
\[
	\til P_{ij} = 1 - Q_i - Q_j + K_{ij} \;.
\]
Then we can build the $S-$matrices 
\begin{equation}\label{eq:fermisma}
	\til S_{ij} = \til P_{ij}\til R_{ij} = 
	1 -2Q_iQ_j +cK_{ij} + (b-1)(Q_i-Q_j)^2\;.
\end{equation}
Unlike in the spin framework, now the relation between 6V $R-$ and
$S-$matrix does not reduce simply to the exchange $c\rightleftharpoons
b$, since Fermi statistics requires that $\til S_{ij}$ must be $-1$ in
the doubly occupied state, rather than $1$. This is taken care by the
last term in eq.(\ref{eq:fermisma}). The matrices $\til S_{ij}$ and
$\til R_{ij}$ manifestly commute with the bare particle number $Q$
which generates the symmetry group $U(1)$. To lightens the notation,
from now on we drop the $\til{}$ throughout, reinstating it only when
strictly necessary.

We have now all the ingredients to build the relevant global objects,
which are the fermionic analog of $V$, $U_R$, $U_L$, $T(\l|\,\{\t_i\})$
and $t(\l|\,\{\t_i\})$, with all the relations that we found in section
(\ref{sec:basic} valid also for the new objects, since they are based solely on
algebraic properties like regularity, YB algebra and permutation
algebra.  In particular, the alternating monodromy matrix
\[
	T =  T(\l|\,\{\t_i\}) = S_{10}S_{20}\ldots S_{2N\,0}
\]
can be written
\[
	 T = A +B\psi_0 + C\psi_0^\da + (D-A)\psi_0^\da\psi_0
\]
where $\psi_0$ and $\psi_0^\da$ are new auxiliary fermion operators
anticommuting with all the previous ones, and $A,\,B,\,C,\,D$ are
global operators in the full fermionic Fock space. Notice that
$\psi_0$ commutes with $A$ and $D$ but anticommutes with $B$ and $C$.
In fact, one easily verifies that $A$ and $D$ have an even fermionic
grade (that is they are sums of terms containing an even number of
$\psi_j$ and $\psi_j^\da$, $j=1,\ldots,2N$), while $B$ and $C$ have an
odd fermionic grade. 

To write the YB algebra it is convenient to rename $\psi_0$ into, say,
$\chi_1$, and introduce another pair $\chi_2$, $\chi_2^\da$,
anticommuting with all $\psi_j$, $j=1,\ldots,2N$ as well as with
$\chi_1$. Then we can write
\[
	 T_r = A +B\chi_r + C\chi_r^\da + (D-A)\chi_r^\da\chi_r
\]
and the YB algebra takes the form of Eq. (\ref{eq:ybalg})
\[
	S_{12}(\l-\mu)\,T_1(\l|\,\{\t_i\})\,T_2(\mu|\,\{\t_i\})=
	T_2(\mu|\,\{\t_i\})\,T_1(\l|\,\{\t_i\})\,S_{12}(\l-\mu)
\]
where (see eq.(\ref{eq:fermisma}))
\[
	S_{12}= 1 + c[\chi_1^\da\chi_2+\chi_2^\da\chi_1]
	+(b-1)(\chi_1^\da\chi_1 + \chi_2^\da\chi_2)
	-2b\chi_1^\da\chi_1\chi_2^\da\chi_2 
\]
To obtain the commutation rules for $A,\,B,\,C,\,D$ the algebra is
now straightforward: one finds some differences of sign with respect
to the rules expressed in eq.(\ref{eq:ybalgI}), namely
\begin{eqnarray}\label{eq:ybalgII}
	b(\mu-\l)A(\l)B(\mu) \!\!&=&\!\! 
	+B(\mu)A(\l)-c(\mu-\l)B(\l)A(\mu)	\nonumber \\
	b(\l-\mu)D(\l)B(\mu) \!\!&=&\!\! 
	-B(\mu)D(\l)+c(\l-\mu)B(\l)D(\mu)	\\
	B(\l)B(\mu) \!\!&=&\!\! -B(\mu) B(\l) 	\nonumber \;.
\end{eqnarray}
Of course, the anticommuting nature of the ``creation operators''
$B(\l)$ appears very natural in this fermionic setup. The other
changes of sign concern only the commutation rules between $D(\l)$ 
and $B(\mu)$, and could be traced to the fact that $S_{ij}=-1$ in the
doubly occupied state.

One last subtlety concerns the meaning of the trace operation. In this
fermionic setup the correct definition would be
\[
	t \equiv  \mathrm{tr}_0\, T = \bra 0 T\ket 0 
	-\bra 1 T\ket 1 = A - D 
\] 
where $\ket 0$ is the state with no auxiliary fermion and $\ket 1$ the
state with one. Then $\mathrm{tr}_0\,P_{j0}=1$ for any $j$ and we find
the fermionic light--cone version of eqs.(\ref{eq:ul}) in the form
\begin{equation}\label{eq:tURULTh}
	t(\Th|\,\Th)  = U_L  \;,\quad 
	t(-\Th|\,\Th) = U_R^\da       \;.
\end{equation}
This choice corresponds to periodic boundary conditions on the
fermions, that is $\psi_{j+2N}\equiv\psi_j$.

On the other hand, we may take as trace what is other contexts is
actually called `supertrace', that is
\[
	t' \equiv \mathrm{str}_0\, T = \bra 0 T\ket 0 
	+\bra 1 T\ket 1 = A + D \;.
\]
Then we find $\mathrm{str}_0\,P_{j0}=1-2Q_j$ and correspondingly (see
eqs.(\ref{eq:trictracI} and eq.(\ref{eq:ul}))
\[
	t'(\Th|\,\Th)  = (1-2Q_{2N}) U_L  \;,\quad 
	t'(-\Th|\,\Th) = U_R ^\da (1-2Q_1)       \;.
\]
In this case we could take the unit--time evolution operator to be
\[
	e^{-ia{\hat H}} = t'(\Th|\,\Th)t'(-\Th|\,\Th)^\da = U'_L U_R
\] 
where as before $U_1=R_{12}R_{34}\ldots R_{2N-1\,2N}$ (see
eq.(\ref{eq:ulur})), while
\begin{eqnarray}
	 U'_L \!\!&=&\!\! (1-2Q_{2N})U_L (1-2Q_1) \nonumber\\
	\!\!&=&\!\! (1-2Q_1) U_2 (1-2Q_{2N}) \nonumber\\ 
	\!\!&=&\!\! R_{23}R_{45}\ldots (1-2Q_1)R_{2N\,1}(1-2Q_1)V \;.
\end{eqnarray}
Similarly we now define the unit--space traslation as
\[
	 e^{iaP} =  t'(\Th|\,\Th)t'(-\Th|\,\Th) = [(1-2Q_{2N})V]^2 \;.
\]
Since $(1-2Q_j)\psi_j(1-2Q_j)=\psi_j$, this choices correspond to
antiperiodic b.c. on the fermion fields
\begin{equation}\label{eq:apbc}
	\psi_{j+2N} \equiv e^{-iLP}\psi_j e^{iLP} = -\psi_j 
\end{equation}
where we have introduced the spatial size of the system $L=Na$.

In summary, we see that for both choices of trace, leading to either
periodic or antiperiodic fermions, as well as in the case of periodic
spins, the nonlocal hamiltonian and total momentum are related to
the tranfer matrix as
\begin{equation}\label{eq:HPandt}
	e^{-ia{\hat H}} = t(\Th|\,\Th)t(-\Th|\,\Th)^\da \;,\quad
	e^{iaP} =  t(\Th|\,\Th)t(-\Th|\,\Th)
\end{equation}
where we may now drop the ${}'$ for the antiperiodic case, provided we
keep in mind the two different ways in which $t(\l|\,\Th)$ is written
in terms of the diagonal elements of the monodromy matrix, either
$A-D$ or $A+D$. It should be clear that identical conclusions about
the the b.c. apply in the framework based on the local hamiltonian of Eq.
(\ref{eq:localham}). 

\sect{Explicit form of the local hamiltonian}\label{sec:form}

We shall now obtain the explicit form of the local hamiltonian $H$ in
terms of the fermionic fields $\psi_j$, $j=1,2,\ldots,2N$. By means of
the Jordan--Wigner transformation one can always revert to the spin
formulation, keeping in mind the effects on the boundary conditions.
For definiteness we shall choose the antiperiodic b.c. for the fermion
fields.  When we insert the expression (\ref{eq:fermisma}) for the 6V
$R-$ matrix into the formula for the hamiltonian density $h(\l)$ (see
eq.(\ref{eq:othercharges})), we need to perform rather long albeit
trivial algebraic manipulations with the fermi fields. In particular
we find
\begin{eqnarray}
	R_{jn}(\l)^\da \dot R_{jn}(\l) \!\!&=&\!\! (\bar b\dot c + 
	\dot b\bar c)K_{jn} +(\bar b\dot b + \dot c\bar c) K_{jn}^2 \\
	R_{jn}(\l)^\da \dot R_{ij}(0) R_{jn}(\l) \!\!&=&\!\!
	\dot b_0 [\psi_i^\da(b\psi_n +c\psi_j) + \mathrm{h.c.}] \nonumber\\
	\!\!&+&\!\! \dot c_0 [b\bar c\psi_j\psi_n +\mathrm{h.c.}) 
	+ Q_i + c\bar c Q_j + b\bar b Q_n] 	\nonumber\\
	\!\!&+&\!\! \dot b_0 [(b+\bar b)Q_jK_{in} + (c-\bar c)Q_n
	(\psi_j^\da\psi_i - \psi_i^\da\psi_j) ]	\nonumber\\
	\!\!&-&\!\! 2\dot c_0 [b\bar c Q_i(\psi_j^\da\psi_n -
	\psi_n^\da\psi_j) + Q_i(c\bar c Q_j + b\bar b Q_n)]
\end{eqnarray}
where $b=b(\l)$, $c=c(\l)$, $\dot b_0=b'(0)$ and $\dot c_0=c'(0)$. In
the derivation of these results the unitarity relations $b\bar b+c\bar
c|=1$ and $b{\bar c}+{\bar b}c=0$ have been used.  To obtain $H$ we
must now set $(i,j,n)=(j-1,j,j+1)$, then put $\l=2\Th$ when $j$ is odd
and $\l=-2\Th$ when $j$ is even, and finally sum up over $j$. $H$
is the sum of a piece quadratic in the fields and a  piece
quartic in them
\begin{eqnarray}\label{eq:explicit}
	H \!\!&=&\!\! H_2 + H_4	 \\
	H_2 \!\!&=&\!\! {{-a_t^{-1}}\o{2\sin\g}}  \sum_{j=1}^N\left[
	d\,\psi^\da_{2j}(\psi_{2j-1}+\psi_{2j+1}) +
	{\bar d}(\psi^\da_{2j-1} + 
	\psi^\da_{2j+1})\psi_{2j}\right. 		\nonumber \\
	\!\!&+&\!\! b (\psi^{\da}_{2j+1} \psi_{2j-1}
	+\psi^{\da}_{2j} \psi_{2j+2}) + {\bar b}(\psi^\da_{2j-1} 
	\psi_{2j+1}+\psi^\da_{2j+2} \psi_{2j}) 		\nonumber \\
	\!\!&+&\!\!  \left. 2 (v+\cos{\gamma})
	(Q_{2j-1}+Q_{2j}) \right]			\nonumber \\
	H_4 \!\!&=&\!\! {{-a_t^{-1}}\o{2\sin\g}}\sum_{j=1}^N 
	\left\{-(b+{\bar b}) (Q_{2j-1}K_{2j-2\,2j} + 
	Q_{2j}K_{2j-1\,2j+1}) \right. 			\nonumber \\
	\!\!&+&\!\!  (c-{\bar c})\left[ Q_{2j} 
	(\psi^\da_{2j-1} \psi_{2j-2} -\psi^\da_{2j-2} \psi_{2j-1})
	+ Q_{2j+1} (\psi^\da_{2j-1}\psi_{2j} - 
	\psi^\da_{2j}\psi_{2j-1} ) \right] 	\nonumber \\
	\!\!&+&\!\!  2i w \cos{\gamma} \left[Q_{2j-2}
	(\psi^\da_{2j-1} \psi_{2j} - \psi^\da_{2j} \psi_{2j-1})
	+ Q_{2j-1} ( \psi^\da_{2j+1} \psi_{2j} - 
	\psi^\da_{2j} \psi_{2j+1}) \right] 	\nonumber\\
	\!\!&-&\!\! \left. 2(v +\cos\g \,{\bar u})
	Q_{2j} \left(Q_{2j-1}+Q_{2j+1}\right) 
	-2 \cos\g \, b{\bar b}(Q_{2j-1}Q_{2j+1}
	+ Q_{2j}Q_{2j+2}) \right\}  		\nonumber
\end{eqnarray}
where 
\begin{eqnarray}
	u \!\!&=&\!\! i\sin\g ({\bar b}{\dot c} + {\dot{b}}{\bar c})
	\nonumber \\  
	v \!\!&=&\!\! i({\bar b}{\dot b}+{\bar c}{\dot c})\nonumber\\
	w \!\!&=&\!\! i b{\bar c} = -i {\bar b}c 	\nonumber\\
	d \!\!&=&\!\! u + i w \cos\g +c			\nonumber\;,
\end{eqnarray}
The quadratic part $H_2$ is better analyzed via the following Fourier
transformation (recall the antiperiodic b.c.)
\begin{equation}\label{eq:campo}
	\pmatrix{\psi_{2j-1} \cr \psi_{2j}} =
	{1\over N} \sum_{-\pi<q\le\pi} 
	\pmatrix{\til\psi_+(q) \cr \til\psi_-(q)} \,e^{iq j} \;,\quad
	q \in {{2\pi}\o N}\left[ \ZZ+ \frac12 \right] \;.
\end{equation}
In the limit $N\to\infty$ of an infinite chain, the sum over $q$
becomes an integral over $q$ running in the first Brillouin zone
$(-\pi,\pi)$. Then $H_2$ takes the form
\[
	H_2 = \int_{-\pi}^\pi dq \,\til\psi(q)^\da h(q)\til\psi(q)
\]
where $h(q)$ is the two--by--two matrix
\[
	h(q) = {{-a_t^{-1}}\o{2\sin\g}} \pmatrix{ 
	be^{-iq}+{\bar b}e^{iq}+2(v+\cos\g) &{\bar d}(1+e^{-iq})\cr
	d(1+e^{iq}) & be^{iq}+{\bar b}e^{-iq}+2(v+\cos\g) \cr} \;.
\]
The two eigenvalues of $h(q)$ represent the bare energy branches of
our lattice model
\begin{equation}\label{eq:energ}
	E_\pm(q)={{-a_t^{-1}}\over{\sin\g}}\left\{2(v+\cos\g)+
	(b+{\bar b})\cos{q}  \pm \zeta [-(b-{\bar b})^2 \sin^2q 
	+ 2d{\bar d}(1+\cos{q})]^{1/2} \right\} 
\end{equation}
where $\zeta$ is $+1$ in the first Brillouin zone and in all the odd
ones, while it is $-1$ in the even zones. These dispersion relations
are depicted in fig.1.
\vskip 1 true cm
	
       \centerline{\epsfig{file=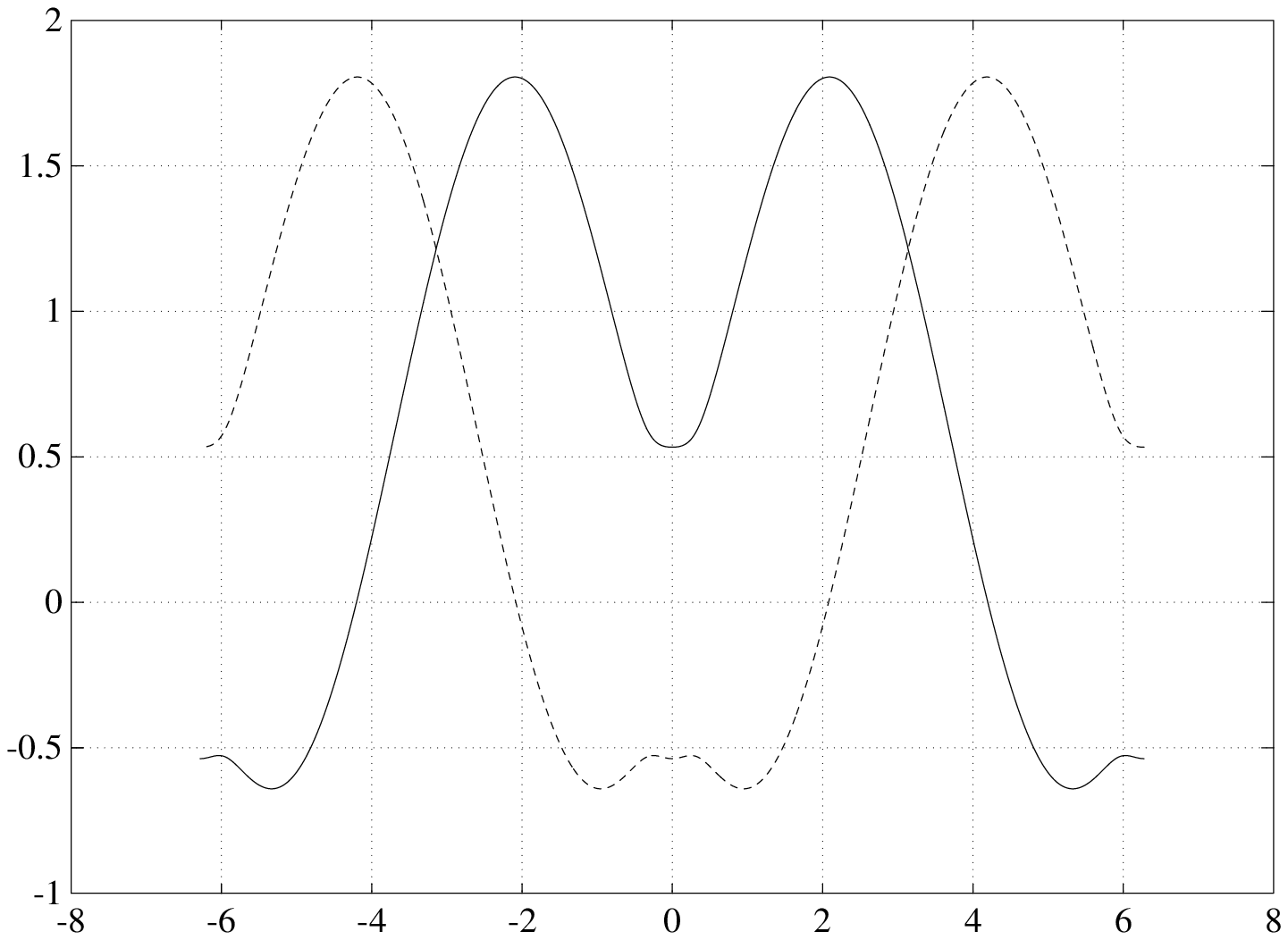,height=5cm}}
 	\centerline{Fig.1: Energy branches for
	$\theta=2$ and $\g=6\pi/10$ in units of $a_t^{-1}$}

\vskip 1 true cm
Evidently all negative energy levels, for both branches within the
first Brillouin zone, should be filled to obtain the lowest energy
state. One must take into account, however, that these bare fermions
are interacting and that this might very well change the shape of the
dispersion relations themselves. The algebraic Bethe Ansatz will take
care of this exactly. At this stage it is enough to assume, as
natural, that in the interaction picture there exist an equal amount
of positive and negative energy levels, so that the perturbative
filled Dirac sea (the perturbative vacuum state of the QFT) is
characterized by half--filling, namely $\VEV{Q_j}=1/2$. This is an
antiferromagnetic state in spin language. It giustifies the
following normal--ordering prescription
\begin{equation}\label{eq:no}
	Q_n= :\!Q_n\!: + {1\over2}
\end{equation}
which has a dramatic effect on the quadratic part of the Hamiltonian,
leading to the perturbative one--particle energy spectrum (see
fig.2)
\begin{equation}\label{eq:energI} 
	E=\pm\frac12\zeta a_t^{-1}{{\sinh{4\Th}}\over{\sinh^2{2\Th} 
	+\sin^2{\gamma}}} \left[\sin^2{q} + 
	{1\over{2}}(m_0a_t)^2 (1+\cos{q})\right]^{1/2}
\end{equation}
where 
\begin{equation}\label{eq:mzero}
	m_0=2a_t^{-1}{{\sin\g}\o{\sinh(2\Th)}} \;\buildrel{\Th\to\infty}
	\over \simeq \; 4a_t^{-1}\sin{\gamma}\,e^{-2\Th} 
\end{equation}
\vskip 1 true cm
 	
	\centerline{\epsfig{file=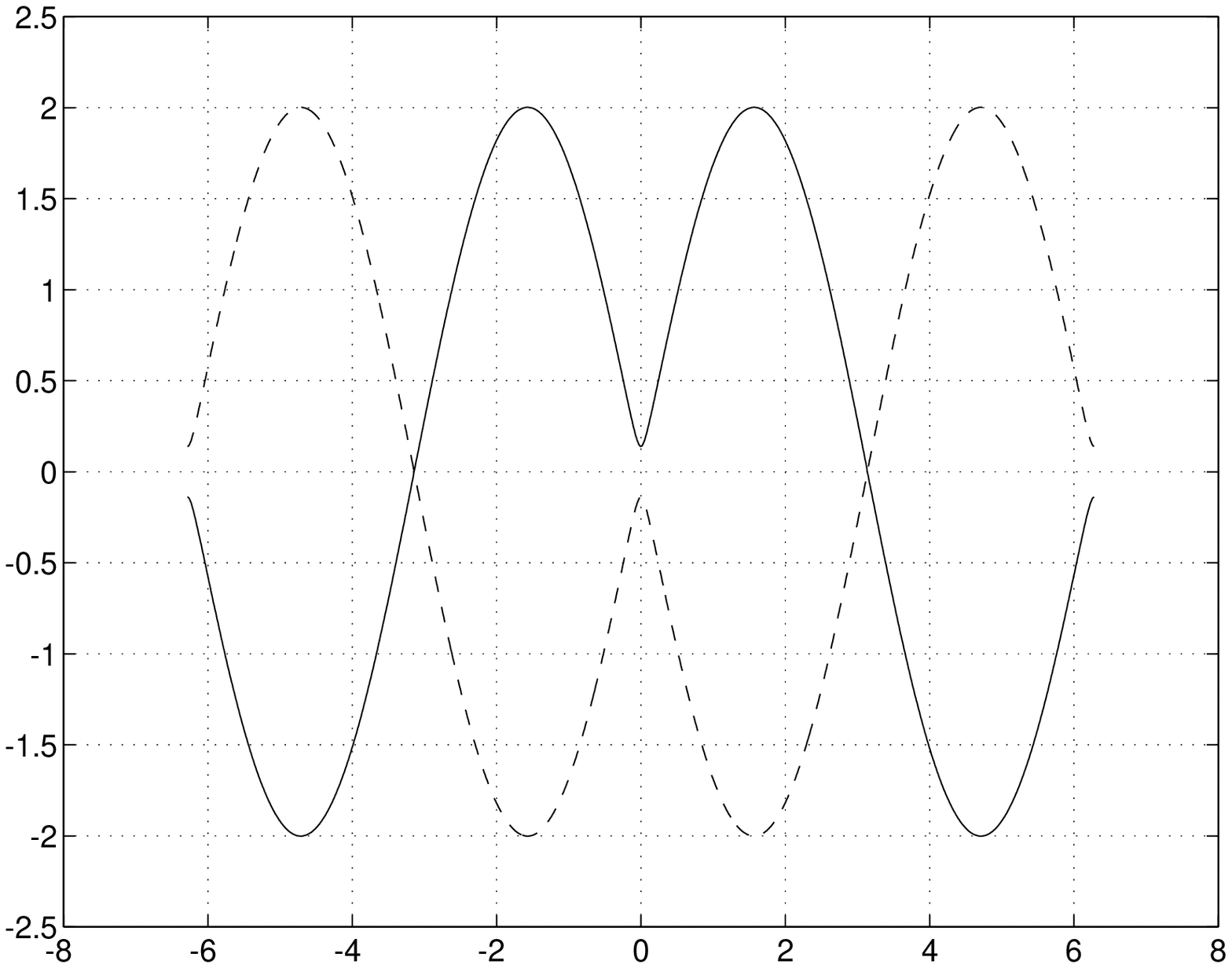,height=5cm}}
 	\centerline{Fig.2: Energy branches after normal--ordering 
	for $\theta=2$ and $\g=6\pi/10$ in units of $a_t^{-1}$}
	
\vskip 1 true cm        
This dispersion relations are manifestly simmetric under reversal of
energy, showing the self--consistency of our normal--ordering
assumption. Once all negative energies in (\ref{eq:energI})
are filled, one obtains a positive spectrum of particles and
holes all with the positive energy of eq.(\ref{eq:energI}). 

It is also clear that eq.(\ref{eq:energI}) represents a lattice
approximation to the relativistic spectrum of massive particles. 
To see this we set $q=pa$ and let $a,\,a_t\to 0$. Then we obtain
\[
	E = c_0 \sqrt{p^2 + m_0^2 c_0^2}
\] 
where $c_0=a/a_t$ is the (bare) velocity of light. It appears
natural to choose spacetime units so that $c_0=1$. Of course one
should expect this choice not to necessarily work in the renormalized
limit to be discussed later. 

The dispersion laws of  eq.(\ref{eq:energI})
has a peculiarity though: it also describes massless particles at the
boundaries of the first Brillouin zone. This is inevitable, since we
are working with a local lattice Hamiltonian which for $\Th\to\infty$,
that is in the massless limit $m_0\to 0$, becomes chiral invariant.
The Nielsen--Ninomiya theorem \cite{nini} then implies the existence
of the (in)famous `fermion doublers'. In the model at hand, the left
and right modes around $q=0$ are massive for finite $\Th$, while
the left and right doubler around $q=\pm\pi$ remain massless. In the limit 
 $\Th\to\infty$ at fixed lattice spacing the model becomes gapless and it
corresponds therefore to a regularized Conformal Field Theory.
According to the general rules, the neighborood of the critical point 
$\Th =\infty$ defines a regularized Perturbed CFT. By letting $\Th \to \infty$
and $a\to 0$ simultanously in a suitable way  one recovers the continuum PCFT. 
The CFT describing the critical point and the perturbing operator will be
identified in the next section.
   
\sect{The continuum limit}\label{sec:continuum}

We now consider the continuum limit $a\to0$ where only the small
energy excitations (as compared with $a^{-1}$) of the fields are
retained and the massless dispersion relations are linearized around
their zeroes \cite{affle3}. 
In this limit the bare mass $m_0$ is kept fixed (the
renormalized continuum limit will be considered in the Bethe ansatz
framework). Thus we must let $\Th\to\infty$ in such a way that 
$m_0\approx 4a^{-1}\sin{\gamma}\,e^{-2\Th}$ stays finite.

The observation of the previous section concerning the doublers
provides the basis for the following representation of the Fermi fields
in the continuum limit 
\begin{eqnarray}\label{eq:deco}                                       
	\psi_{2j} &\!\!\simeq&\!\! \sqrt{a}\left( \chi_L(ja) + 
		(-)^j \eta_R(ja)\right) \nonumber\\
	\psi_{2j\!+\!1} &\!\!\simeq &\!\! \sqrt{a}\left( \chi_R(ja) + 
			(-)^j \eta_L(ja)\right) 
\end{eqnarray}   
where $\chi$ and $\eta$ are quantum relativistic Dirac fields. The
hopping operator $K_{i,i+2}$ and the local charge operator
$Q_i$, e.g. for $i$ even, read
\begin{eqnarray}
	K_{i,i+2} &\!\!\simeq&\!\! 2a :\!\chi^\da_L\chi_L\!:-
	:\!\eta^\da_R\eta_R\!:		\nonumber\\
        :\!Q_{2j}\!:  &\!\!\simeq&\!\! a[:\!\chi^\da_L\chi_L\!: 
	+ :\!\eta^\da_R \eta_R\!: -(-)^j 
	(\chi^\da_R\eta_L+\eta^\da_L\chi_R)] \;.\nonumber
\end{eqnarray}	
The symbol $:\ldots:$ on the r.h.s. refers to the usual
normal--ordering for continuum fields in the interaction picture. This
holds because the operators on the l.h.s. have vanishing vacuum
expectation value. The complementary cases, namely $Q_{2j-1}$ and
$K_{2j-1,2j+1}$, can be handled analogously simply by exchanging right
and left modes. These are all the calculations needed to obtain the
continuum limit of $H_4$, since all quartic terms except the first and
the last are suppressed as $a\to0$. As for the quadratic piece
$H_2$, the typical calculation reads
\begin{eqnarray}
	\psi^\da_{2j}\psi_{2j+2} \simeq a\left[\chi^\da_L(x)\chi_L(x+a)
	- \eta^\da_R(x)\eta_R(x+a)\right]	\nonumber\\
	+ (-)^j\left[\eta^\da_R(x)\chi_L(x+a)
	-\chi^\da_L(x)\eta_R(x+a)\right]	\;,	\nonumber
\end{eqnarray}
so that, dropping the oscillating terms and developping to first
order in $a$ we can calculate the non-vanishing terms in the quadratic
Hamiltonian as
\[
	\psi^\da_{2j}\psi_{2j+2} -\psi^\da_{2j+2}\psi_{2j} \simeq
	2a (\chi^\da_L \partial_x\chi_L-\eta^\da_R\partial_x\eta_R)\;.
\]
Thus, taking into account the normal--ordering and dropping as above
all fast oscillating terms, the continuum form of the Hamiltonian
reads (here $\Th\to\infty$ as $a\to0$ so that the bare mass $m_0$
stays fixed)
\begin{equation}\label{eq:Hcont}
	H \longrightarrow 
		H_0 + H_{\mathrm{m}} + H_{\mathrm{int}}
\end{equation}
where $H_0$ is the kinetic energy 
\begin{equation}\label{eq:h0}
	H_0 = -i \int dx \left(
	\chi^\da_R \partial_x\chi_R -
	\chi^\da_L \partial_x\chi_L + 
	\eta^\da_R \partial_x\eta_R -
	\eta^\da_L \partial_x\eta_L  \right) \;,
\end{equation}
$H_{\mathrm{m}}$ is the mass term 
\begin{equation}\label{eq:hm}
	H_{\mathrm{m}} = m_0  \int dx \left( 
	\chi^\da_L\chi_R+ \chi^\da_R\chi_L \right)
\end{equation}
and $H_{\mathrm{int}}$ the quartic interaction
\begin{equation}\label{eq:hint}
	H_{\mathrm{int}} = 2g' \int dx \left(
	J^\chi_R  J^\chi_L -  J^\eta_R  J^\eta_L -
	J^\eta_R  J^\chi_L -  J^\chi_R  J^\eta_L  \right)  \;.
\end{equation}
Here $g'=-2\cot\g$ 
and the $J$'s are free--field normal--ordered $U(1)$ currents:
\[
	J^\chi_\a = \,:\!\chi_\a^\da \chi_\a \!: \;,\quad
	J^\eta_\a = \,:\!\eta_\a^\da \eta_\a \!: \;,\quad \a=R,L\;.
\]
The nice feature of this result is that all terms surviving the
na\"{\i}ve continuum limit are manifestly Lorentz--invariant, unlike those
obtained in the analogous treatment of the XXZ spin chain in \cite{affle3}.
It is natural to regard the mass term $H_m$ as `bare'
perturbation of the CFT defined by $H_0 +H_{int}$.
The troublesome aspect is that $\eta$, the field describing the doublers,
does not decouples from the putative massive Thirring field $\chi$ and prevents
a straightforward identification of the CFT.

In order to find  the right decoupled description, we use
abelian bosonization:
\begin{equation}\label{eq:boso}
	\matrix{
	\chi_\a = \mu^{1/2} :\!\exp(i\a\sqrt{4\pi}\,u_\a)\!:\cr
	\eta_\a = \mu^{1/2} :\!\exp(i\a\sqrt{4\pi}\,v_\a)\!:  \cr}
\end{equation}
where $\a=\pm$ ($+\equiv R$ and $-\equiv L$), $\mu$ is a normalization
mass scale, and the fields
\[
	u_\a(x) = \int_{-\infty}^x dy\,J^\chi_\a(y) \;,\quad
	v^\eta_\a(x) = \int_{-\infty}^x dy\,J^\eta_\a(y)
\]
can be identified with the chiral components of two free massless Bose
fields. The symbols $:\ldots:$ in eq.(\ref{eq:boso}) now stand for
bosonic free--field normal ordering at the mass scale $\mu$, so that
the expressions (\ref{eq:boso}) are effectively $\mu$-independent.

With the standard rules of abelian bosonization, the Hamiltonian
now takes the form, up to irrelevant constants,
\begin{eqnarray}
	H = \int dx \!\!\!\!\!\!&&\!\!\!\!\!\!\left\{
	(\partial_x u_R )^2 + (\partial_x u_L )^2 +
	(\partial_x v_R )^2 + (\partial_x v_L )^2  \right. 
	\nonumber \\ \!\! &+&\!\! g' \left[ 
	(\partial_x u_R)(\partial_x u_L) -
	(\partial_x v_R)(\partial_x v_L) -
	(\partial_x u_R)(\partial_x v_L) -
	(\partial_x v_R)(\partial_x u_L) \right]  \nonumber \\
	\!\!&+&\!\! \left. m_0\mu:\!\cos[\sqrt{4\pi}(u_R+u_L)]\!:
	\right\} 
\end{eqnarray}
Notice that only one boson field is involved in the sine--Gordon
interaction, but the mixed terms in the third line still couple the
two boson fields. In order to elimate them we can use the canonical
transformations that leave invariant  the commmutation rules between
left and right components of the boson fields:
\begin{eqnarray}
	\left[u_{\a}(x),u_{\a'}(x')\right] \!\!&=&\!\! 
	\frac{i}4 \a \delta_{\a\a'} \epsilon(x-x') \nonumber\\
 	\left[v_{\a}(x),v_{\a'}(x')\right] \!\!&=&\!\!
	\frac{i}4 \a \delta_{\a\a'} \epsilon(x-x') \nonumber\\
	\left[u_{\a}(x),v_{\a'}(x')\right] \!\!&=&\!\!  0 \;,  
\end{eqnarray}              
Thus it must be a $O(2,2)$ trasformation.  We find it combining two
canonical $U(1,1)$ transformation and a canonical ortogonal
$SO(2)\times SO(2)$ transformation acting on right and left sectors
separately.  We obtain in this way:
\begin{equation}\label{eq:transftot}
	\left(\matrix{  \phi_R\cr
              		\xi_R\cr   
              		\phi_L\cr
              		\xi_L	}\right) =
	\left(\matrix{	 r & t &-s & -t\cr
              		 t &-r & t & -s\cr
              		-s &-t & r &  t\cr
             		 t &-s & t & -r	}\right)
	\left(\matrix{   u_R \cr
              		 v_R \cr
             	 	 u_L \cr
              		 v_L 	}\right)
\end{equation}
with
\begin{eqnarray}
	r = {{\cosh^2\nu \cosh\l +\sinh^2\nu\sinh\l}
		\over{\sqrt{\cosh 2\nu}}} 	\nonumber \\
	s = {{\cosh^2\nu \sinh\l +\sinh^2\nu\cosh\l}
		\over{\sqrt{\cosh 2\nu}}} 	\nonumber \\
	t = {{\cosh\nu\sinh\nu \left( \cosh\l -\sinh\l\right)}
		\over{\sqrt{\cosh 2\nu}}}	\nonumber            
\end{eqnarray}
and
\[
	\tanh(2\l)= -{g'\over{2\pi}}  \;,\quad 
	\tanh(2\nu)= -\sinh(2\l) 	\;.
\]
In terms of the new fields the Hamiltonian reads
\begin{eqnarray}\label{eq:sine}
	H &=&\left[1-{{g'^2}\over{2\pi^2}}\right]^{1/2}\int dx\left\{
	(\partial_x \phi_R )^2 + (\partial_x \phi_L )^2 +
	(\partial_x \xi_R )^2 + (\partial_x \xi_L )^2 \right\} 
	\nonumber \\  &+&  
	m_0\mu \int dx \,:\!\cos[\sqrt{4\pi}\,e^\l(\phi_R+\phi_L)]\!:
\end{eqnarray}
and we see that it correspond to a sine--Gordon model plus a decoupled
free massless field.
More precisely, in passing to lagrangian form, we should scale the
fields $\phi$ and $\xi$ so that the kinetic term is properly
normalized. In this way one arrives at the Lagrangian:
\begin{equation}\label{eq:sinegordon}
	{\cal L}=\frac{1}{2}(\partial_\mu\phi)^2 +
	\frac{1}{2}(\partial_\mu\xi)^2 + m_0\mu\cos\beta\phi 
\end{equation} 
where $\phi=\phi_L+\phi_R$ and $\xi=\xi_L+\xi_R$. 
The relation of Coleman's coupling constant $\b$ with $g'$ reads  
\begin{equation}\label{eq:betag'}
	{{\beta^2}\over{4\pi}}={{1+g'/\pi}\o{1-2(g'/\pi)^2}}\;.
\end{equation} 
We could now perform the inverse bosonization trick on $\phi$ and
$\xi$, according to the standard rules \cite{abdalla}, or with
canonical transformation analogous to those done above. This yields
at the end two decoupled Thirring models, one massive, with Dirac
field $\psi$, and one massless, with field $\psi'$:
\begin{equation}\label{eq:thirring}
	{\cal L} =\bar{\psi}(i\gamma^\mu\partial_\mu -m_0)\psi 
	+\frac{1}{2}g(\bar{\psi}\gamma^\mu\psi)^2+
  	\bar{\psi'}(i\gamma^\mu\partial_\mu)\psi'+
	\frac{1}{2}g(\bar{\psi'}\gamma^\mu\psi')^2 \;.
\end{equation}
We see therefore the fermion doubling, charateristic of any local
lattice regularization with local chiral currents, is completely
harmless in our case: it only adds a decoupled massless field to the 
Lagrangian.  

Our derivation is now complete: we have shown that the fermion
Hamiltonian (\ref{eq:explicit}) provides a local lattice
regularization of the massive Thirring model. The important point is
that this Hamiltonian is completely integrable, being just the first
of an infinite hierarchy of conserved charges in involution. One may
regard all terms in the lattice Hamiltonian which are of order $a$ as
irrelevant operators needed to preserve the integrability on the
lattice.

Of course we have performed a `bare' continuum limit which does not
take into account renormalization effects. However, the integrability
of the model allows to include them exactly through the explicit
diagonalization of the lattice Hamiltonian. This is carried through by
means of the algebraic Bethe ansatz, or Quantum Inverse Scattering
Method, whose main steps will be outlined in the next section.

On the basis of such Bethe ansatz, on--shell solution, the
simultaneous presence of massive and massless particles in the
relativistic QFT describing the continuum limit was already put
forward in ref.\cite{rs}.  Here we have related this fact to the
well--known phenomenon of fermion doubling and have in particular
clarified its non trivial off--shell extension.

\sect{Main results of the Bethe ansatz}\label{sec:BA}

By definition, the {\em algebraic} Bethe ansatz will work in the
fermionic formulation just like in the standard spin framework. All
changes of sign due to the fermionic commutation rules
(\ref{eq:ybalgII}) can be easily traced down.  Wee need not repeat here
any derivation, referring to the various review articles on the
subject (see for instance \cite{devega}).

The eigenvectors of the alternating tranfer matrix are written
(see eq.(\ref{eq:mono}))
\begin{equation}\label{eq:baeigvTh}
 	\ket\Psi = B(\l_1+i\g/2|\,\Th)B(\l_2+i\g/2|\,\Th) \ldots 
		B(\l_r+i\g/2|\,\Th)\, \ket\Om
\end{equation}
where $\ket\Om$ is the bare vacuum state and the parameters
$\l_1,\l_2,\ldots,\l_r$ satisfy the Bethe ansatz equations (BAE)
\begin{equation}\label{eq:baeTh}
	\left[ {{\sinh(\l_m + \Th + i\g/2)} \o 
		{\sinh(\l_m + \Th - i\g/2)}}\right]^N   
	\left[ {{\sinh(\l_m - \Th + i\g/2)} \o
		{\sinh(\l_m - \Th - i\g/2)}}\right]^N 
	= (-)^{r+1} \prod_{n=1}^r {{\sinh(\l_m -\l_n +i\g)}
		 \o {\sinh(\l_m-\l_n -i\g)}} \;.
\end{equation}
The eigenvalues read
\begin{eqnarray}\label{eq:lamdasTh}
	\La   \!\!&=&\!\! \La^A + \La^D		\nonumber\\
	\La^A \!\!&=&\!\! \prod_{k=1}^r {{\sinh(i\g/2 + \l - \l_k)} 
				     \o {\sinh(i\g/2 - \l + \l_k)}}\\
	\La^D \!\!&=&\!\! (-)^r b(\l+\Th)^N \,b(\l-\Th)^N 
	\prod_{k=1}^r	{{\sinh(3i\g/2 - \l + \l_k)} 
		      \o {\sinh(-i\g/2 + \l - \l_k)}} \;.
\end{eqnarray}
Since one easily verifies that $[Q\,,\,B]=-B$, the BA states
(\ref{eq:baeigvTh}) contain exactly $r$ bare particles. Notice also that
eigenvectors and eigenvalues depend on $\Th$ both explicitly and
through the dependence forced on the numbers $\l_k$ by the BAE. We do
not need to consider states with more than $N$ bare particles, since
they are obtained by particle--hole symmetry, {\it i.e.} $\psi_j
\rightleftharpoons \psi_j^\da$, corresponding to spin inversion in
spin language, from the states (\ref{eq:baeigvTh}).

As usual, we introduce the so--called {\it counting function}
\cite{devega}
\[
	Z_N(\l)= N\left[\phi(\l+\Th,\g/2)+\phi(\l-\Th,\g/2) \right] -
	\sum_{k=1}^r \phi(\l-\l_k,\g)			
\]
where
\[
	\phi(\l,x)\equiv i\log{{\sinh(ix+\l)}\o{\sinh(ix-\l)}}
\]
has the cut structure chosen so that it is analytic in the strip
$|\Im\l| \le x$. The BAE now read 
\[
	Z_N(\l_j) = 2\pi I_j \;,\quad j=1,2,\ldots,r
\]
where the quantum numbers $I_j$ are always half--odd--integers 
(we choose $N$ to be even). This should be compared with the spin
formulation where the $I_j$ are half--odd--integers for even $r$ and 
integers for odd $r$. This appears very natural if we compare the b.c.
of antiperiodic fermion fields, eq.(\ref{eq:apbc}), with those
corresponding to periodic spins, eq.(\ref{eq:Ftbc}). This difference
will play a crucial role in determining the $U(1)$ charge of the
physical particles. 

The energy (both local and nonlocal) and momentum of a given BA state
are calculated from eqs.(\ref{eq:charges}), (\ref{eq:localham}),
(\ref{eq:HPandt}) and (\ref{eq:lamdasTh}). The momentum reads 
\begin{equation}\label{eq:prpart}
	P = a^{-1} \sum_{j=1}^r \left[\phi(\Th+\l_j,\g/2) 
	- \phi(\Th-\l_j,\g/2) \right]
\end{equation}
while the local energy is  
\begin{equation}\label{eq:erpart}
	E = -\frac12 a_t^{-1}\sum_{j=1}^r 
	\left[\der{}{\Th}\phi(\Th+\l_j,\g/2) +
		\der{}{\Th}\phi(\Th-\l_j,\g/2)\right]\;.
\end{equation}
The physical vacuum state, or filled Dirac sea, is the ground state of
the local Hamiltonian $H$, that is the lowest possible value of $E$
for fixed $N$. It corresponds to the unique solution of
the BAE with $N$ real roots. In the limit $N\to\infty$ at fixed
lattice spacing $a$ (hence in the infinite volume limit), this
solution is described by a smooth density \cite{devega}. The same
applies to all particle states characterized by a finite number of holes
in the ground state distribution. The energy and momentum of 
one of this holes (a physical fermion) can be calculated
exactly to be
\begin{equation}\label{eq:ep1part}
   E(\varphi)= \frac12 a_t^{-1}  \der{p(\varphi)}{\varphi}  \;,\quad
   p(\varphi)= 2{a\o{a_t}}\arctan\left({{\sinh{\pi\varphi/\g}}\o
            {\cosh{\pi\Th/\g}}}\right)                      
\end{equation}
where $\varphi$ is the position of the hole in the Dirac sea.
This is all rather standard. The important novelty concerns 
the $U(1)$ charge of the holes.

In the usual spin formulation with periodic b.c., to the removal of a
single BA root there corresponds the appearence of {\em two} holes.
Therefore each hole has a renormalized charge $Q=-1/2$. This is
clearly incompatible with the interpretation of such holes as
fermions, since they would not be interpolated by the fermi fields
$\psi_n$.  The sign differences proper of the fermionic framework, and
in particular the factor $(-1)^r$ in eq.(\ref{eq:baeTh}), exactly remedy
this. An accurate analysis of the phase space available for $N-1$ BA
roots, using the asymptotic value of the counting function $Z_N(\l)$,
shows that only one hole is present in the Dirac sea. This is the
dressed antiparticle of the original fermion and has charge $Q=-1$. As
a matter of fact one can consider also states with $N+1$ BA roots, one
of which has imaginary part equal to $i\pi/2$: one finds the same
energy--momentum spectrum of eq.(\ref{eq:ep1part}), while evidently
$Q=1$. The dressed particle is obtained by particle--hole
symmetry. 

The states with one particle and one hole are obtained by
removing one real BA root and introducing a root with immaginary part
equal to $\pi/2$.This naturally follows by looking at the dependence
of energy and momentum on the ``lattice rapidities'' $\l_j$ in
eqs.(\ref{eq:prpart}) and (\ref{eq:erpart}): the replacement $\l_j\to
\l_j+i\pi/2$ exchanges the two energy branches in eq.(\ref{eq:energ}).
 
It is possible to identify the solutions of the BAE corresponding to
states with arbitrary many fermions and antifermions as well as with
breathers (fermion-antifermion bound states in the attractive regime
$\g>\pi/2$). A complete and detailed analysis is still lacking in the
literature (parts can be found in the early BA approaches to the
continuum massive Thirring model \cite{kore}\cite{bergo} and in the
general study of the BA equations for the XXZ chain \cite{babe}), but
is outside the scopes of this work.

For our purposes, it is enough here to examine the continuum limit of
the massive part of the renormalized energy--momentum.  As $a\,,a_t\to
0$ and $\Th \to\infty$ we find from eqs.(\ref{eq:ep1part}) the
relativistic expressions in terms of the rapidity
$\t=\displaystyle{{\pi\varphi}\o\g}$:
\begin{equation}\label{eq:ep1particella}
	E = mc^2 \cosh\t  \;,\quad  p = mc \sinh\t 
\end{equation}
provided we identify the renormalized velocity of light and
mass as:
\begin{equation}\label{eq:exact}
	c = {{\pi a}\o{2\g a_t}} \;,\quad
	mc = {{4}\o{a_t}} e^{-\pi\Th/\g} \;.
\end{equation}
Notice that the velocity of light undergoes a finite renormalization
from the bare value $c_0=a/a_t$ found before. Of course, we could set
the conventional $c=1$ by adjusting $a/a_t$ to $2\g/\pi$. Notice also
that eliminating $\varphi$ from the two relations in
(\ref{eq:ep1part}) one obtains, still on the lattice
\begin{equation}\label{eq:energIren} 
	E=  a^{-1}c \,\tanh\left(\frac{\pi\Th}{\g}\right)
	 \left[\sin^2{pa} + \frac12(mc\,a_t)^2 (1+\cos{pa})\right]^{1/2}
\end{equation}
with the more precise mass definition
$mc\,a_t=2/\sinh(\pi\Th/\g)$. This renormalized dispersion relation
should be compared with the perturbative one, eq.(\ref{eq:energI}):
apart from an overall factor which tends to $1$ as $\Th\to\infty$, all
renormalization effects are concentrated in the rescalings $m_0\to m$
and $c_0\to c$. In particular the exact spectrum (\ref{eq:energIren})
has the same fermion doublers of the perturbative one: such doublers
are still coupled to the massive modes, as could be checked with the
direct calculation of the relevant scattering phase shifts. In the
continuum limits the characteristic momenta of the massless and
massive modes get separated by a quantity of order $a^{-1}$ and these
scattering phase shifts tend to non--trivial constants. The off--shell
decoupling shown in the previous section ensures that a proper
additional dressing of the massive and massless particles exists that
decouples them altogether.

\sect{ On the relation between the coupling costants }\label{sec:gg'b}

We may now investigate more in details the connection between the
parameters of the lattice Hamiltonian and those of the continuum ones,
either bosonic (sine--Gordon) or fermionic (massive Thirring).  

The standard relation between $\b$ in eq.(\ref{eq:sinegordon}) and the
coupling constant $g$ in eq.(\ref{eq:thirring}) reads
(see {\em e.g.} \cite{cole}\cite{abdalla})
\begin{equation}\label{eq:betag}
	{{\beta^2}\over{4\pi}}={1\over{1-g/\pi}}
\end{equation} 
and differs from the relation (\ref{eq:betag'}) derived above
with $g'$. The sine--Gordon coupling constant $\b$ is a
regularization--independent parameter, since, for $\b^2<8\pi^2$, the
sine--Gordon model can be uniquely defined as a perturbated conformal
theory \cite{eguchi}.  Hence we may safely take $\b$ as a reference
parameter to relate $g $ and $g'$:
\begin{equation}\label{eq:gg'}
	g=g' {{1+2g'/\pi}\o{1+g'/\pi}}=g'\left(1+{{g'/\pi}\o{1+g'/\pi}}
	\right) = g'\left[1+\sum_{n=1}^{\infty} (-1)^{n-1}
	\left({{g'}\o\pi}\right)^n \right] \;.
\end{equation} 
They differ by a formal power series redefinition, as to be expected
in the Thirring model, since the current--current coupling in two
dimensions is cutoff independent but regularization--scheme dependent.

Let us observe, moreover, that the relation (\ref{eq:gg'}) holds in the
interaction picture, since we are applying to the interacting
sine--Gordon field theory the bosonization rules proper of the free
bose field. We can relate more precisely $\b$ to the well--defined
lattice parameter $\g$, and then to $g'=-2\cot\g$ (see
eq.(\ref{eq:hint})), by using exact scaling arguments as follows.  The
scaling dimension of $\cos\b\phi$ is $\b^2/4\pi$, since it is fixed by
the ultraviolet fixed point, namely the free massless bose field.
Through bosonisation $\cos\b\phi$ maps into ${\bar\psi}\psi$, which
enters the lagrangian of the massive Thirring model multiplied by
$m_0$. Hence $m_0$ has scaling dimension $2-\b^2/4\pi$. From the
exact Bethe ansatz solution one learns that the physical mass scale 
is proportional to $\exp(-\pi\Th/\g)$ (see eq.(\ref{eq:exact})). 
On the other hand, eq.(\ref{eq:mzero}) shows that
$m_0$ scales like $\exp(-2\Th)$, so that it has scale dimension
$2\g/\pi$. Therefore we must have $2\g/\pi=2-\b^2/4\pi$, which is the
exact relation we sought.

This argument is quick but rather too sketchy. A more precise
derivation goes at follows. The redefinitions of the normalization
mass scale $\mu$ and those of the the bare mass $m_0$ are connected by
the normal--ordering renormalization group \cite{cole}, in order to
keep $m_0{\bar\psi}\psi=m_0\mu:\!\cos\b\phi\!:$ invariant. This leads
to the relation
\[
	{{m_0\mu}\o{m_0'\mu'}} = \left({\mu \o{\mu'}}\right)^\Delta 
\]
where $\Delta=\b^2/4\pi$. On dimensional grounds, the physical mass
scale has the form
\[
	m = m_0 f(\b,z) \;,\quad z = {{m_0}\o\mu} 
\]
and must be renormalization--group invariant, that is 
\[
	m = m_0 \l^{\Delta+1} f(\b, z\l^{\Delta+2}) 
	\;,\quad \l = {\mu \o{\mu'}} \;.
\]
Hence $f$ is a homogeneous function of $z$ and we obtain
\begin{equation}\label{eq:mphys}
	m = m_0 z^{-y}\, f(\b,1) = m_0^{1-y}\mu^y  f(\b,1) 
\end{equation} 
where $y={{\Delta+1}\o{\Delta+2}}$.

The exact Bethe ansatz solution of the lattice model provides the
following relation for the fermion mass in the continuum limit
\[
	m \simeq 4 {{a_t}\o{a^2}} {{2\g}\o{\pi}} e^{-\pi\Th/g} 
	={{2\g}\o{\pi}}\left({{a_t}\o{a}}\right)^2 \left({{m_0}\o{\sin\g}}
	\right)^{\pi/2\g} \left({a_t\o4}\right)^{\pi/2\g-1} 
\]
where eq.(\ref{eq:mzero}) was used in the second equality.
Choosing $\mu=a_t^{-1}$ and $a/a_t=2\g/\pi$, to enforce $c=1$, we
obtain, comparing to eq.(\ref{eq:mphys}),
\[
	{{\b^2}\o{8\pi}} = 1 - {\g\o\pi} \;,\quad f(\b,1) = 
	{{16\pi}\o{8\pi-\b^2}}
	\,\left[4\sin(\b^2/8)\right]^{4\pi/(\b^2-8\pi)} \;.
\]
The relation between $\g$ and $\b$ is that we found above. In addition
we found an expression for $f(\b,1)$. Of course this expression is
scheme--dependent.

Coming back to the Thirring coupling constants $g$ and $g'$, we have
the following situation: $g$ is defined through bosonization of the
massive Thirring model alone and is given by $g=\pi-4\pi^2/\b^2$ (see
eq.(\ref{eq:betag})). Hence we have the exact relation
\begin{equation}\label{eq:gammag}
	g = {\pi\o2}\, {{\pi-2\g}\o{\pi-\g}} \;.
\end{equation}
$g'$ may be analogously defined through bosonization of the complete
continuum hamiltonian (\ref{eq:Hcont}) (which contains the fermion
doublers). This leads to the exact relation (\ref{eq:gg'}).  On the
other hand we have the relation $g'=-2\cot\g$ (see below
eq.(\ref{eq:hint})), which follows from the continuum limit of the
lattice hamiltonian in the interaction picture (using free--field
normal--ordering), and therefore is only approximate or ``bare''.
Combining eqs.(\ref{eq:gg'}) and (\ref{eq:gammag}) we obtain the exact
relation
\[
	{{2g'}\o\pi} {{1+2g'/\pi}\o{1+g'/\pi}} = 
	{{1-2\g/\pi}\o{1-\g/\pi}}
\]
which can be regarded as the renormalization of the bare relation
$g'=-2\cot\g$.

\sect{Final comments and outlook}\label{sec:fine}

The local lattice regularization of the massive Thirring model
presented here applies equally well to the vast class of integrable
models already under control by means of the standard light--cone
approach. The local character of the lattice Hamiltonian should help
in the field--theoretic understanding of these models, since it allows
for a better control of the continuum limit and an easier
identification of each model as a perturbed CFT. From this
field--theoretic point of view, the most important step remains the
proper definition of the local lattice fields in terms of which the
$R$--matrices are to be written. When this is done, the Hamiltonian as
well as all other conserved charge, either local or nonlocal, would
follow by the standard techniques of vertex models, since only the
algebraic properties of the $R$--matrices and the permutation
operators are needed. In the case of the massive Thirring models this
program hass been here pursued explicitly starting from the local
$R$--matrices written in terms of canonical lattice fermi fields
(eq.\ref{eq:fermirma}) and handling the continuum limit as in section
\ref{sec:continuum}.

\end{document}